\documentclass[floats,amsfonts,amssymb,unsortedaddress,floatfix]{revtex4}
\input epsf
\usepackage{bm}
\usepackage{amsmath}
\usepackage{graphicx}
\usepackage{xcolor}
\usepackage{ulem}

\begin{document}

\title{More Opportunities than Wealth: A Network of Power and Frustration}

\author{$^*$Benoit Mahault,
$^+$Avadh Saxena and $^+$Cristiano Nisoli}

\affiliation{$^*$Service de Physique de l'Etat Condens\'e, CNRS UMR 3680, CEA-Saclay, 91191 Gif-sur-Yvette, France \\
\mbox{$^+$Theoretical Division and Center for NonLinear Studies, Los Alamos National Laboratory, Los Alamos NM 87545 USA}}

\begin{abstract}
{We introduce a minimal agent-based model to qualitatively conceptualize the allocation of limited wealth among more abundant opportunities. We  study the interplay of power, satisfaction and frustration in    distribution, concentration, and inequality of wealth. Our framework allows us  to compare subjective measures of frustration and satisfaction to collective measures of fairness in wealth distribution, such as the Lorenz curve and the Gini index. We find that a completely libertarian, law-of-the-jungle setting, where every agent can acquire wealth from, or lose wealth to, anybody else invariably leads to a complete polarization of the distribution of wealth vs. opportunity. 
The picture is however dramatically modified when hard constraints are imposed over agents, and they are limited to share wealth with neighbors on a network.  
We then propose an out of equilibrium  dynamics {\it of} the networks, based on a competition between power and frustration in the decision-making of agents that leads to network coevolution. We show that the ratio of power and frustration controls different dynamical regimes separated by kinetic transitions and characterized by drastically different values of the indices of equality. The interplay of power and frustration leads to the emergence of three self-organized social classes, lower, middle, and upper class, whose interactions drive a cyclical regime.
}
\end{abstract}

\maketitle


\section{Introduction}
Since Vilfredo Pareto provided the earliest quantitative analyses of wealth distribution~\cite{Pareto1896,Pareto1897new}, the inequality of societies  has been a matter of debate and concern  for economists, social scientists, and politicians alike. Its correlation, anti-correlation, or lack thereof, with growth is historically among the most debated issues in political economics~\cite{Marx,benabou1996inequality,persky1992retrospectives,kuznets1955economic}. Wealth inequality has recently caused growing concerns for the functioning of the democratic institutions~\cite{gilens2014testing,gilens2012affluence,bonica2013hasn} and occupied the  political debate~\cite{krugman2013inequality}. 

While convincing econometric studies have reported increasing polarization of {\it income} in western societies~\cite{deininger1996new,mishel2012state},  more  recent  data document a worldwide increase in {\it wealth} inequality~\cite{mishel2012state,saez2014wealth,piketty2014capital,jones2015pareto,piketty2014inequality}.   In general  ``wealth is unequally distributed, more so than wages or incomes''~\cite{mishel2012state} and ``Gini coefficients for wealth typically lie in the range of about 0.6-0.8. In contrast, most Gini coefficients for disposable income fall in the range 0.3-0.5''~\cite{davies2011level}. The Gini index~\cite{gini1912}  for the wealth of the entire world (and indeed of the USA) is estimated at a dramatic 0.8 with the bottom 50\% of the world population owning 3.7\% of wealth (ref. \cite{davies2011level} and references therein)~\footnote{The Gini coefficient is an index of statistical dispersion for a distribution introduced by Corrado Gini~\cite{gini1912} to describe wealth or income inequality. It is computed from a Lorenz curve, which plots cumulative  wealth vs. cumulative population. The Gini index is the normalized area that lies between the line of equality $y=x$ and the Lorenz curve: it is 0 for perfect equality and 1 for perfect inequality. See Fig.~2 and Fig.~4 for examples.}. 

It has been generally understood that wealth inequality follows income inequality, the delay corresponding to accumulation of savings among the upper-income strata~\cite{Pareto1896,demirguc2009finance}. This process can be long  (ref. \cite{jones2015pareto} and references therein) impinging  on issues of inherited wealth and social stratification~\cite{spilerman2000wealth}. However, faster, more direct pathways to wealth polarization might be possible in the fluid contemporary world where increasingly deregulated  and sophisticated financial tools can provide new means for wealth transfer~\cite{palley2007financialization}. 

Simple examples abound. Much of the middle class' wealth is stored in real estate, the acquisition of which involves financial instruments  that effectively deflect more than half of the painfully saved wealth towards other agents (or indeed all of it, when epidemic foreclosures follow a  bubble inflated by financial deregulation~\cite{stephens2007mortgage,andrews2008greenspan}). Also,  in the aftermath of a market crisis the least wealthy are at loss and cannot capitalize on 
on the following market rebound  as the wealthy can, providing a ratchet mechanism toward wealth inequality at each significant market oscillation.  And indeed  wealth inequality in the USA worsened  after the 2007 financial crisis even though income inequality  mildly ameliorated~\cite{wolff2013asset}. Finally, as deregulated financial instruments offset wage stagnation and the widening gap between productivity and living standards~\cite{mishel2012state},   a steady increase in household debt to finance consumption further contributes to wealth polarization~\cite{mishel2012state,barba2009rising}.  Indeed, the USA, at the forefront of financial deregulation,  also tops most countries in wealth inequality, with a Gini index of $\sim 0.8$ (worse than any African country except Namibia, as of year 2000)~\cite{davies2011level}. Furthermore,  the American trend inversion concerning the share of the top 1\% ~\cite{saez2014wealth} began in the late 70s, closely tracking the deregulation of the financial market~\cite{sherman2009short}.

\begin{figure*}[t!]
\centerline{\includegraphics[width=.98\textwidth]{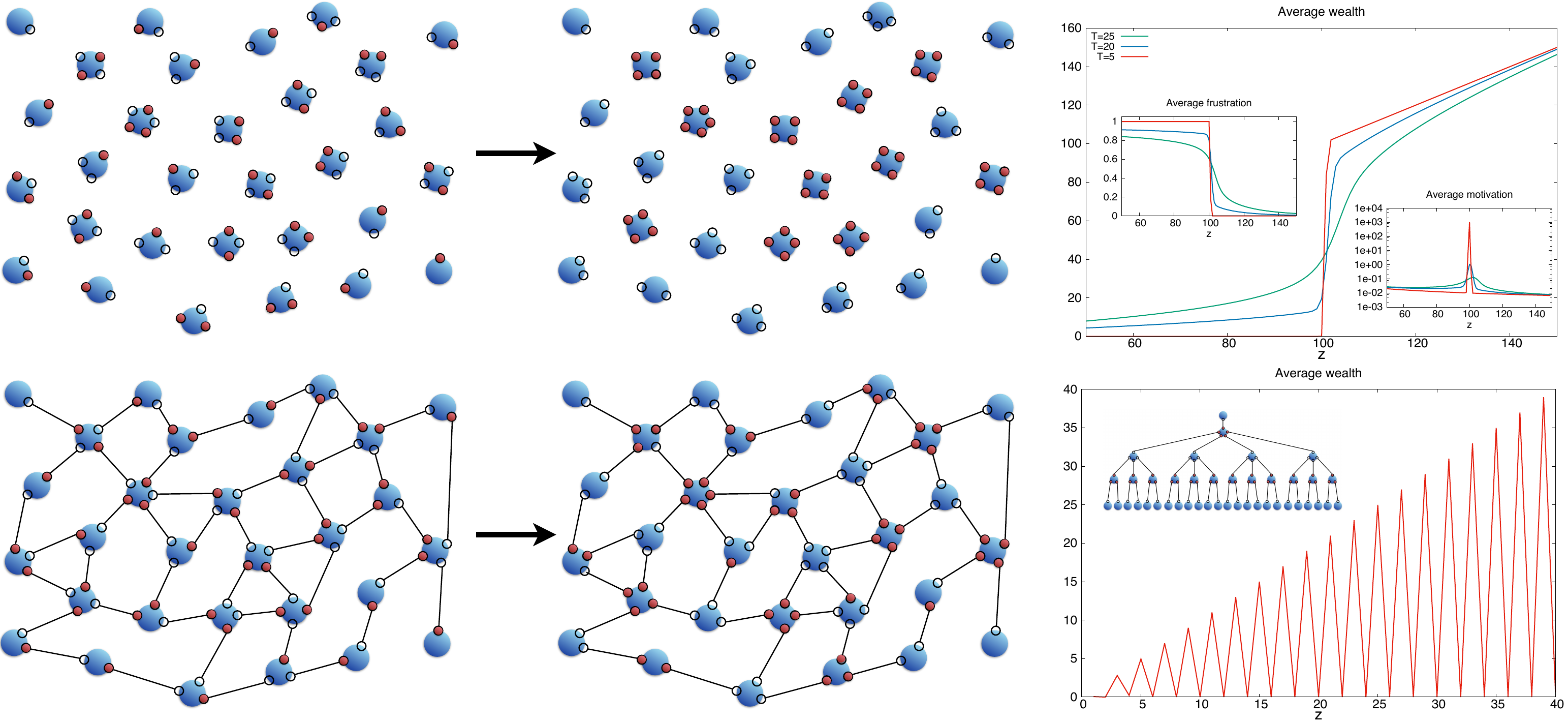}}
\caption{Schematic of the model. Top left: The "Law of the jungle" where every agent (in blue) can grab wealth (in red) from any other to fill its available opportunities (circles). The market power is optimized by a polarized society where agents with most opportunities get all their opportunities satisfied. Top right: The distribution of average wealth $\langle w_z\rangle$ vs. opportunities $z$ at different temperatures in the law of the jungle setting for a binomial distribution of opportunities of mean $\bar z=100$. The left and right insets plot respectively the average frustration and average motivation vs. opportunities $z$. Bottom left: same society as above but with hard constraints, which limit exchanges to pairwise action among agents and its optimized state. Bottom right: The distribution of average wealth vs. opportunity optimizing power for a hierarchical fractal tree with $l=40$ (The inset reports the case $l=5$) as a (pathological) example of how hard constraints can severely frustrate wealth polarization.}
\label{afoto}
\end{figure*}

While the issue is certainly very much complex, it motivates a conceptualization of  direct pathways to wealth accumulation/polarization, in a framework reflecting  the essential features: power, inequality, frustration, and initiative.

In this work we explore how wealth inequality  appears naturally in a  minimal model of agents endowed with opportunities to grab or lose fractions of  limited available wealth. The approach is inspired by our research in  physics, chemistry and statistical mechanics~\cite{Nisoli2013colloquium,Nisoli2014NJoP, Zhang2013, Wang2006,gilbert2014emergent} which  provide rigorous means to predict statistical equilibria, in contrast to general equilibria  in economics, and to also follow frustrated dynamics, which is relevant for inequality. 
In particular, our framework allows us to compare  subjective (frustration, satisfaction) as well as global (Lorenz curve, Gini index) measures of fairness, and study their interplay in the evolution of society. 

\section{Results and Discussion}

\subsection{The Law of the Jungle}
Consider a population of agents all endowed with varying number of opportunities $z$. These are defined as available slots, which  might or might not be filled by  $w$ units of wealth, as in Fig.~1.  $P_z$ is the normalized distribution of opportunities and represents the market.  There are twice as many opportunities as there is wealth ($\bar w= \bar z/2$ where $\bar w$ and $\bar z$ are the average wealth and opportunities): half of all the opportunities go frustrated. We can quantify the {\it individual frustration $f$} and {\it satisfaction $s$} of an agent with $z$ opportunities who has gathered wealth $w$ as
\begin{align}
& s=w/z \\
& f=1-w/z
 \label{frus}
 \end{align}
 where $f=1$, $s=0$ means complete frustration (no satisfied opportunities), $f=0$, $s=1$ complete satisfaction (all opportunities are satisfied)~\footnote{Strictly speaking, it is redundant to use both these terms, as they
are trivially related to each other. But, depending on the situation, one or the other seems much
more natural.}.

We assume that the {\it power} of wealth to attract more wealth  does not scale linearly with gathered wealth. We define the power ${\cal P}$ of an agent as interaction between its wealth, or
\begin{equation}
{\cal P}=w(w-1)/2,
\end{equation}
and look for wealth distributions that maximize the total power in the market, as the sum of the power of its agents. We will study $\langle w_z \rangle$ the average wealth of agents of opportunity $z$, as well as its fluctuations, or $\sqrt{\langle w_z^2 \rangle}$. From 
it we  compute the social motivation 
$m_z=\left({\langle w_{z+1} \rangle-\langle w_z \rangle}\right)/{\langle w_z \rangle}
\label{mot}
$, 
 a collective property   describing the marginal increase in wealth for an increase in opportunities. 

From $P_z$ and $\langle w_z \rangle$ we  compute the Lorenz curve of cumulative wealth vs. cumulative population and  the Gini index  $G_w$ to assess the global fairness of a society. We  also introduce average {\it individual} measures of fairness to describe the return on opportunity, such as average frustration $\bar f =\sum_z \langle f_z \rangle P_z$ and average satisfaction $\bar s$. Of those we consider the Lorenz curve {\it for personal satisfaction} rather than wealth and the corresponding Gini index $G_s$. 
$G_s=0$ implies that all agents have the same satisfaction (and frustration) equal to 1/2--a generalization of what physicists call the ``ice rule''~\cite{Pauling1935,Ramirez1999,Nisoli2013colloquium,Nisoli2014NJoP, Zhang2013, Wang2006,gilbert2014emergent,Libal2006}. 


The distribution of wealth which maximizes the total power of the market can be found easily by filling  with wealth  the agents of largest  opportunities and continuing in decreasing order of opportunity, until all  wealth is stored. The result is a completely polarized society in which agents with opportunity above a critical $z_c$ have zero frustration and all their opportunities satisfied, whereas the others have nothing~\footnote{If $P_{z_c}\ne0$  half of the agents with $z_c$ opportunities will have all opportunities satisfied, the other half will have none.
} as in Fig.~1. There is  no ``middle class''. The critical opportunity $z_c$ is implicitly defined by
\begin{equation}
\sum_{z<z_c}zP_z =\frac{\overline{z}}{2}. 
\label{zcrit}
\end{equation}
Since  $ P_zz /\bar z$  is the distribution of maximal expectations  vs. opportunities, all the expectations below the median fall in the class of the have-nots. The fraction of the utterly dispossessed depends on the structure of the market, or $P_z$, but it is always larger than 50\% of the population~\footnote{It is exactly 50\% when all the agents have  the same number of opportunities~\cite{Nisoli2014NJoP}: even then, half of the population gets nothing.}. 
In particular, if the average wealth is large enough~\footnote{A realistic assumption which we will always make, as the unit of transaction must be small compared with the average wealth.} the fraction of dispossessed is scale-invariant in the opportunities, and recipes often touted by politicians, such as ``increasing opportunities for everybody''  are not viable policies in absence of a structural change of the society. Indeed for large $\bar w$, we can safely take the continuum limit on $z$, and see that a  scaling  $P_z \to kP_{kz}$ does not change  the overall fraction of have-nots, as the critical opportunity also scales in Eq.~(\ref{zcrit}). 

One way to grow a small middle class is to make society less efficient for power. We can introduce ``thermal'' disorder into the system via standard statistical physics techniques~\cite{Nisoli2014NJoP} already applied to social settings~\cite{dragulescu2000statistical,yakovenko2009colloquium}, where a parameter $T$, an effective temperature,  describes the deviation of the market from optimal behavior. Fig.~1 shows how the middle class is characterized at low $T$  by intermediate frustration and is the most motivated. As  society becomes more optimized for power ($T\to0$), the  middle class shrinks to nothing, while its motivation skyrockets~\footnote{ At very large disorder  powers play no role and the ice-rule wins with $\langle f_z \rangle=1/2$ leading to an egalitarian society, but that is a trivial and not very realistic case.}. 


\subsection{Constrained Societies}

Until the mid $20^{th}$ century, individuals of large opportunities could not grab wealth from just anybody else, as in the  model above. The owner of a company was mostly linked to his workers---and could subtract from them part of the wealth they produced in the form of added value---yet was not similarly linked to other agents. We consider thus the effect on wealth inequality of limiting wealth transfer to pairwise transactions. This is achieved by connecting agents as nodes of a network of shared edges. On each edge now sits the unit of wealth, and it can be grabbed or lost by the two nodes, as in Fig. 1. The number of opportunities of an agent corresponds to node coordination and $P_z$ is now the degree distribution of the graph.  
As network theory has been employed successfully in recent years to describe a number of complex systems, from the World Wide Web, to financial markets, neuronal connections, and the growth of cities~\cite{cohen2010complex,dorogovtsev2008critical,gao2012networks,Barabasi1999,albert2002statistical,bettencourt2007growth,bettencourt2013origins,amaral2000classes}, we apply it here to describe wealth transfer. 

An example demonstrates how  a network setting can radically modify the previous  scenario:  consider the graph of Fig.~1, a hierarchical  tree where the agent with most opportunities $l$ is connected to agents with opportunities $l-1$ and so on. In the limit of large $l$ we have a self-similar fractal and the distribution of wealth can be solved by iteration (see SI). As Fig.~1 shows, the relationship between wealth and opportunities for this network is not even monotonic, precisely because each agent is connected to agents of similar opportunities.

We  explore  dynamics {\it on} the networks   via a Metropolis Monte-Carlo algorithm. Each step corresponds to the transfer of a single unit of wealth among two randomly chosen connected agents, and is accepted if it increases the power, or else with probability exponentially decreasing in the decrease of power (${\cal P}'-{\cal P}$),  proportional to $\exp[ ({\cal P}'-{\cal P})/T]$. The degenerate manifold of maximum power is obtained when  $T\to 0$. We will confine ourselves to small values of $T$ typically of order less than 1. 
We find that the implicit mean field formula
\begin{equation}
\langle w_z\rangle=z\sum_{z'}\frac{P(z'|z)}{1+ \exp \left[(\langle w_{z^{'}}\rangle-\langle w_z\rangle)/T\right]}
\label{constraintw}
\end{equation}
 excellently fits the numerical results (see SI for a  derivation). Here $P(z'|z)$, is the conditional probability that an agent of opportunity $z$ has agents of opportunity $z'$ among its connected partners. Equation~(\ref{constraintw}) can be easily understood: the average wealth of agents is proportional to their opportunity $z$ and to the fraction of connected partners of coordination $z'$, weighted with a mean thermal factor describing the probability of acquiring the unit of wealth.
 
For an optimized society ($T \to 0$) Eq.~(\ref{constraintw}) returns~\footnote{Under the assumption that wealth increases monotonically with opportunities.}   the average satisfaction as
%
$\langle s_z\rangle=P(z|z)/2+\sum_{z'<z}P(z'|z)$,
%
or  the fraction of neighbors with equal or lower opportunities; satisfaction is maximal ($s=1$) only when all neighbors have fewer opportunities, which lends credence to long held  popular wisdom. 


\begin{figure*}[t]
\begin{center}
\centerline{\includegraphics[width=.9\textwidth]{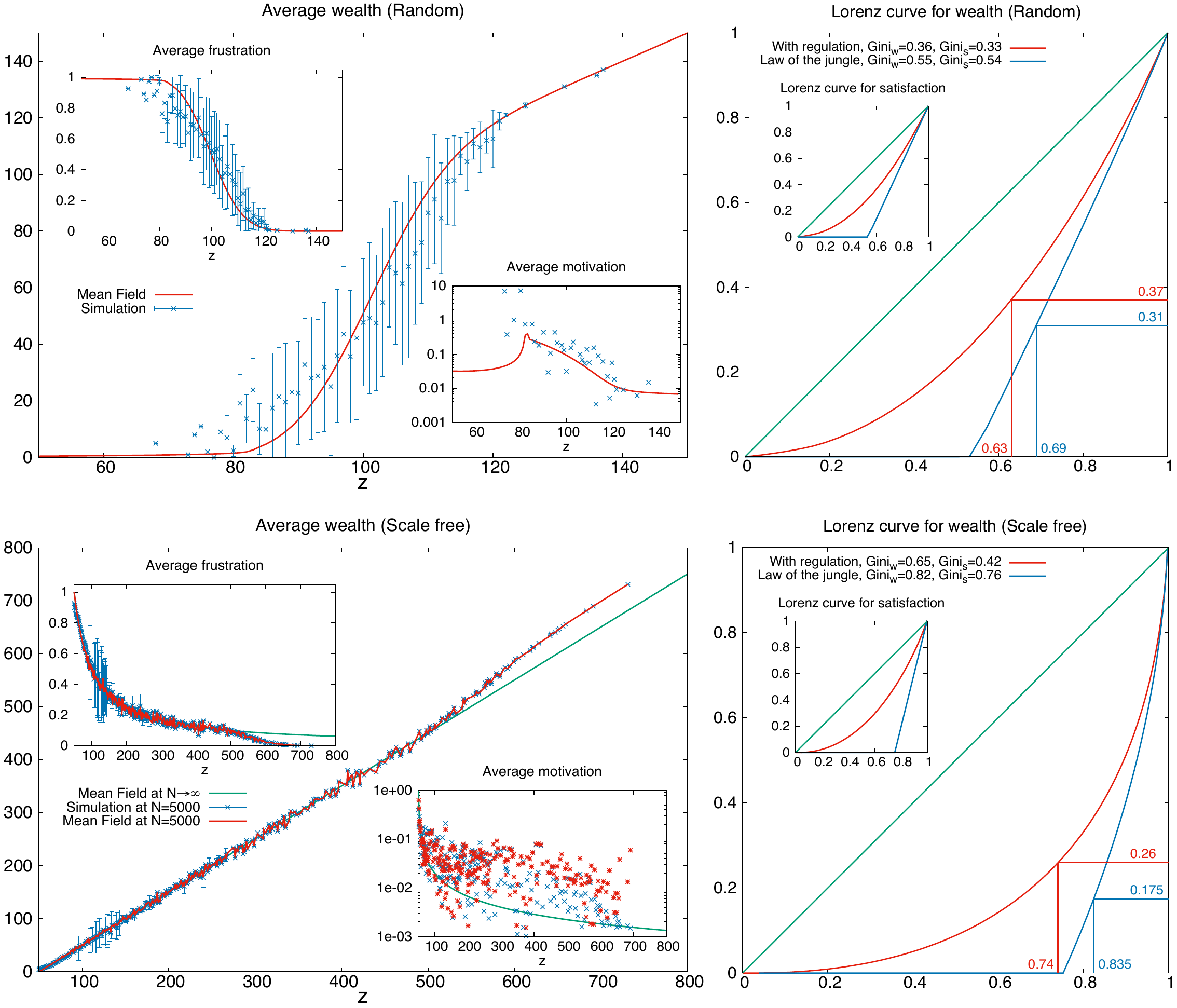}}
\caption{Results on random and scale free networks. Top: The distribution of average wealth $\langle w_z\rangle$ vs. opportunities $z$ for a random network with $N=1000$ nodes and $p=0.1$ probability of connection (thus average wealth $\bar{w}=50$) at $T=1$, obtained via Metropolis simulation (in red the mean field formula). The left (right) inset represents the distribution of average frustration $\langle f_z\rangle$  (motivation $m_z$) vs. opportunities $z$. The bars denote fluctuations in wealth (and frustration) among agents endowed with the same opportunities due to the topological structure of the graph. On the right we represent the Lorenz curves for cumulative  wealth  (in inset satisfaction) vs. cumulative population,  their corresponding Gini coefficients, and Pareto squares  with and without hard constraints. in  power-optimized societies. Bottom: Same figures for a Barab\`asi-Albert graph with minimum coordination $m=\bar{w}=50$ in a finite system ($N=5000$ nodes)  in the thermodynamic limit at $T=0.1$. (The green line describes the wealth allocation  obtained by using $P(z'|z)$  for the scale free network (reported in SI) in the limit of a large society. Here finite size effects cause a deviation of  $\sim5\times10^{-4}$  agents among the richest. The red line is obtained via the mean field treatment using the {\it actual} conditional probability of the finite system and reproduces the numerical results remarkably well.)}
\label{afoto2}
\end{center}
\end{figure*}

\subsection{Random Networks, Scale Free Networks}

We first test our framework on  random graphs and scale free networks. The former were until recently assumed to be the most natural and common form of networks. The latter, however, have been more recently proposed to describe e.g. the Internet~\cite{Barabasi1999},    sexual encounters~\cite{liljeros2003sexual} and alliance  in industry~\cite{powell1996interorganizational}. 

For random networks  we employ the   Erd\H os-R\'enyi model~\cite{Erdos1961} where  
%
 $N$ is the number of vertices, $p$ the probability of connection, and the degree distribution is binomial (see SI for details). 
 The average opportunity is $\bar z=pN$ and in the thermodynamic limit of large $N$ the degree distribution is Poisson if the average wealth $\bar w=\bar z/2=pN/2$ is kept constant.

We also build scale free networks   via the Barab\'asi-Albert preferential attachment algorithm~\cite{Barabasi1999} which has a scale free degree distribution  $P_z~\sim z^{-3}$ at large $z$,  corresponding to highly correlated nodes called hubs.  
Choosing the minimal coordination $m$, we then have $\bar z=2m$ (see SI), and the average wealth is $\bar w=m$: unlike in the  Erd\H os-R\'enyi model, all agents possess enough opportunities to own the average wealth. One would thus expect this network to be fairer. It is  not. 

Fig.~2 reports Monte-Carlo simulations for these two networks along with results for a law-of-the-jungle society of corresponding distribution of opportunities. The allocation of wealth vs. opportunities is quite different in the two cases. 

In the random graph case one can distinguish three social classes, a lower class of very high frustration, a middle class of average opportunities and of frustration  centered around $1/2$,  and an upper class of large opportunity and zero frustration.
However due to the Poisson distribution of opportunities, most agents belong to the middle class and both the Lorenz curve and the low Gini index for wealth ($G_w\simeq0.36$) and personal satisfaction ($G_s\simeq0.33$) reflect this fairness. Furthermore, the middle class is  characterized by the highest motivation and very large fluctuations of  wealth for given opportunity, something that Pareto already noted in real societies~\cite{Pareto1896,Pareto1897new}. Using our mean field formula we see that in the limit of a large society the average frustration is $\bar f=1/2$.

Yet the same fair distribution of opportunities leads to a dramatically different scenario  in a law-of-the-jungle case: while the Gini index is still decent ($G_w\simeq G_s\simeq 0.5$) and lower than the US Gini, in this case it proves to be a poor indicator of fairness: a look at the Lorenz curve shows that more than half of the population is completely dispossessed and  unsatisfied. 

In the scale free case, both numerics and our mean field treatment returns, remarkably, a linear dependence of wealth vs. opportunities, or  $\langle w_z\rangle\simeq z-\bar w$.
Thus no clear class distinction can be deduced in Fig.~2 and the average motivation simply increases as opportunities decrease as $m_z\simeq1/(z-\bar w)$. In such a network at least the average wealth is accessible to everybody:  everybody has relatively more opportunities than in a random network. Yet we see a  worse Lorenz curve and  Gini index ($G_w\simeq0.65$) for wealth. Interestingly, however, while collectively less fair, the network appears, in a more subjective way, reasonably fair: the Lorenz curve and Gini index for the individual  are not that much different from the previous case: the society is more polarized in plain wealth but not so much in the return on opportunities; the opportunities are simply distributed differently, with a few ``hubs'' getting most of them~\cite{Barabasi1999}. 

Here too a law-of-the-jungle setting returns a dramatic  profile of inequality with 75\% of the population completely dispossessed and dissatisfied. By contrast, in the connected case, the fraction of the completely dispossessed goes to zero  in the realistic limit of large average wealth $\bar w$.  Indeed, it corresponds to agents of   lowest opportunity $z=\bar w$, whose fraction is  $2/(\bar w+2)$~\cite{fotouhi2013degree}.

The fairness of the Random network  is due to its narrower, symmetric distribution of opportunities. As we will see, when the system is allowed to evolve out of equilibrium at the interplay of power and frustration, it will often try and sometimes succeed in recreating such situation, typically by producing interdependent random networks.

\begin{figure}[t]
\begin{center}
\centerline{\includegraphics[width=.9\columnwidth]{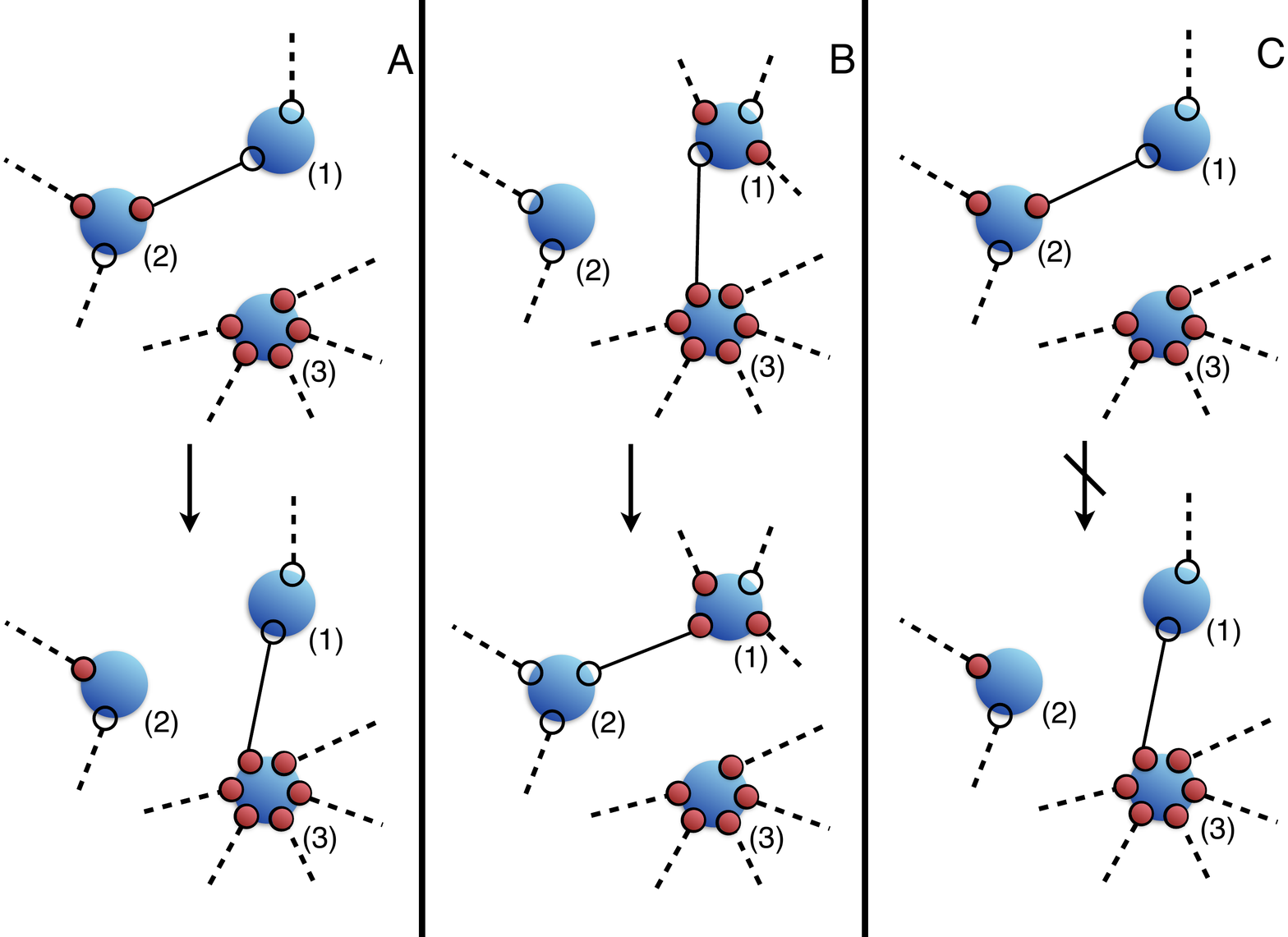}}
\caption{Dynamics of social interactions. Move A is promoted by power alone: (3) is much richer than (1) and (2) thus it will take the quantum of wealth they share. Move B and C  on the contrary are promoted by frustration as in Eq.~(\ref{Move}). In B (3) is not frustrated but (1) and (2) are, thus  the probability for them to reorganize their neighborhood is increased. Move C is  suppressed by the social friction implicit in Eq.~(\ref{Move}).}
\label{afoto2}
\end{center}
\end{figure}

\subsection{Emergent Social Classes and Kinetic Transitions in Market Evolution}

Societies and markets are not static and seldom in  equilibrium. We can extend our framework by allowing agents to make decisions on the basis of their status to take, if not wealth, at least opportunities from others, or to break ties with neighbors and establish new relationships  thus leading to an evolution--indeed coevolution~\cite{gross2008adaptive,barrat2008dynamical}--of market topology. At the simplest level of modeling, our previous analysis on static markets suggests that such moves can be dictated by sheer power and also personal initiative to act on frustration. 

As the total wealth is constant, for any broken partnership a new one must emerge. In general the transition $\{a-b,c\}\to \{a-c,b\}$  involves three randomly grouped agents, $a,b,c$: $a$ breaks ties with $b$ and joins $c$, $b$ loses an opportunity, $c$ gains one, although this does not necessarily translate in gain or loss of actual wealth. Risk is involved.   

If the dynamics is based on power alone, then $c$ will gain wealth as well as an opportunity if it is more powerful than $a$. This is depicted in transition A of Fig.~3, which is promoted by a power-only dynamics where a move is accepted with a probability proportional to $\exp[ ({\cal P}'-{\cal P})/T]$ where ${\cal P}'-{\cal P}$ is the difference in power through the move. Simulations show that at low disorder $T$, this process rapidly reaches an expected equilibrium,  the best market for power: a Gini index close to 1 (Fig.~4), complete inequality, and very few agents with extremely large opportunities, subtracting all the wealth from the remaining, low-coordinated network.  
 
\begin{figure*}[t!]
\begin{center}
\centerline{\includegraphics[width=\textwidth]{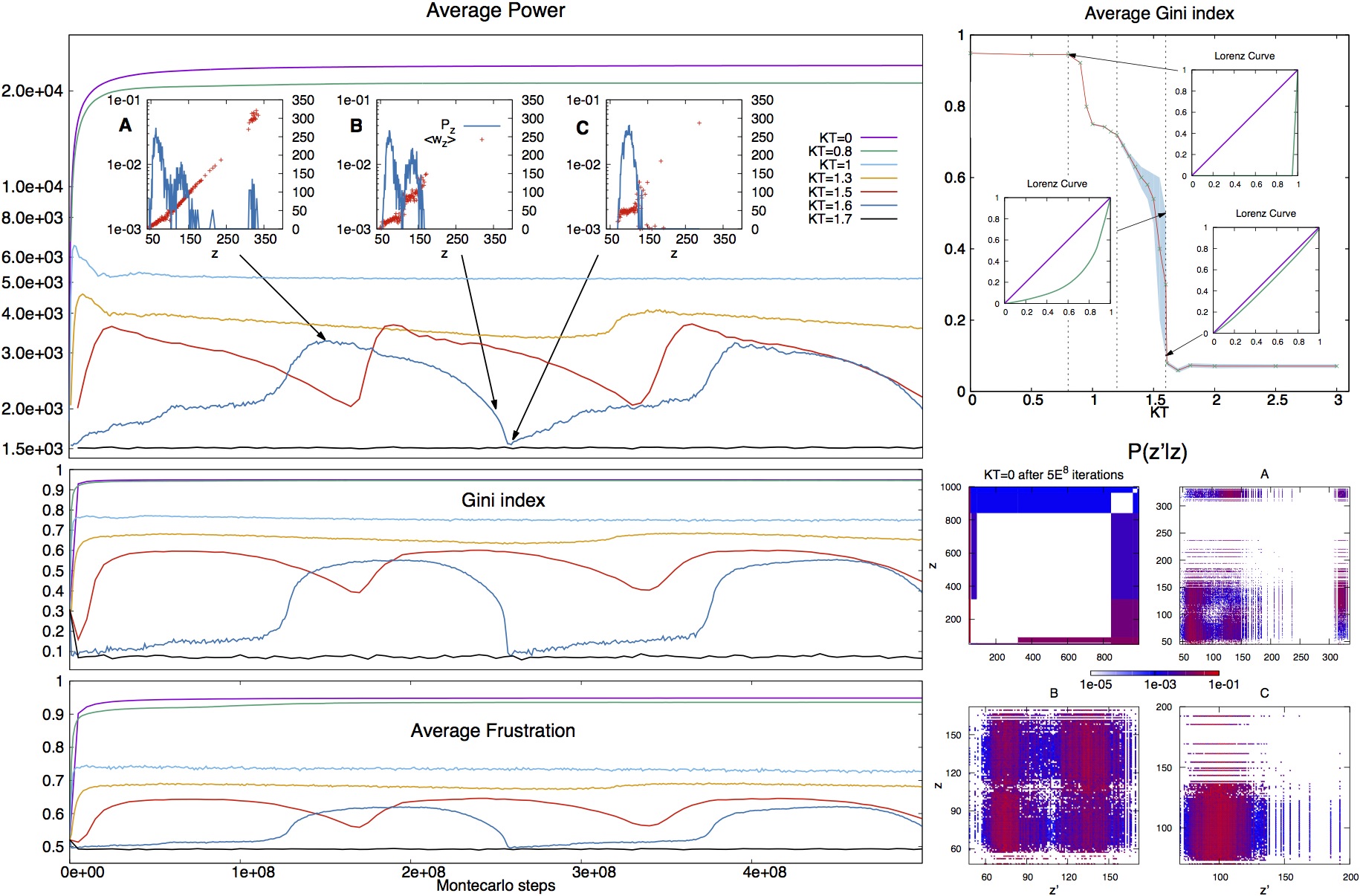}}
\caption{Effect of initiative on the coevolution of society. Left panels: Average power, Gini index and average frustration at different values of initiative $K$  evolving in time (Monte Carlo steps),  starting from an equilibrated random graph with $N=1000$ nodes and $p=0.1$ attachment probability (thus average wealth $\bar w=50$, constant during the evolution), at minimal disorder $T=0.1$. The A,B and C insets represent the average wealth (red, values on right axis) and degree distributions (blue, values on left axis) for $KT=1.6$ at times $1.5\cdot10^8$ (three social classes), $2.5\cdot10^8$ (two classes) and $2.6\cdot10^8$ (one class). Top right: The mean Gini index as a function of $KT$, and its kinetic transitions (the blue surface corresponds to the periodic regime). The insets show the Lorenz curves at ($KT=0.8$,$G=0.94$), ($KT=1.6$,$G=0.55$) and ($KT=1.6$,$G=0.09$). The dashed lines distinguish four kinetic regimes: from left to right the power-dominated, the softened power, the periodic, and the egalitarian regimes. Bottom right: The conditional degree distribution of the graphs for insets A, B, C and also in the $KT=0$ case after $5\cdot 10^8$ iterations, showing that in the absolute-power regime all of the most coordinated agents are principally connected to the lower class.}\label{afoto2}
\end{center}
\end{figure*}

However, in a more realistic setting each agent is motivated to act on his/her frustration and also, if frustrated enough, to resist the action of the powerful. We can express that by adding a new kind of probability for a step. In the general case the probability to accept a step $\{a-b,c\}\to \{a-c,b\}$, where $a$ is the active agent, $b$ is the old partner, and $c$ is the new partner, is proportional to the product of the previous, power-maximizing factor $\exp[ ({\cal P}'-{\cal P})/T]$ times the extra factor 
\begin{align}
\exp\{K[(w_b-w_c)f_a+(w_c-w_a)f_c-w_af_b]\},
\label{Move}
\end{align}
where we call $K$  ``initiative'' and all the wealths and frustrations of the agents are taken {\it in the initial configuration}. Equation~(\ref{Move}) is the product of three probabilities implying that a frustrated agent is more likely to change partner  if the new partner is less wealthy than the old one. The more the new partner is frustrated, the more it is willing to acquire the new opportunity, but only if sharing wealth with a less wealthy partner.  The old partner always resists losing an opportunity. 

These three intentions, however, combine themselves to allow or suppress different transitions. For instance, in  Fig.~3 we see that  transition B where two frustrated agents establish a new connection at the expense of a more powerful and less frustrated partner, is promoted. Agent (1) will be motivated to shift partnership away from agent (3) which is less frustrated and more powerful. Moreover, as (3) is much richer than (2) and un-frustrated, the greatly frustrated (2) will be motivated to accept the association with (1) even though the move  proves not fortunate in terms of wealth.  The factor in Eq.~(\ref{Move}) also supresses  the transition C of Fig.~3  because (1) and (2) are still frustrated and this move would have a negative effect on (1) who has fewer chances to get the unit of wealth at the end. 

We note that the extra term due to frustration represents an out of equilibrium drive, as it is based on the initial configuration, rather than on the difference of configurations through the step: each move  carries risk, there is no assurance that a switch in partnership will bring wealth to the agent. (SI provides a table of  Eq.~(\ref{Move}) for extreme cases.)

Figure~4 and the animation in SI show results for simulations at different values of $K$ at fixed $T=0.1$, revealing intriguing kinetic transitions~\footnote{The initial state is a random network, however, using a scale free network does not change the picture (see SI and animations 2 and 3 for the same kinetics applied to both networks). We believe this process is independent of the initial condition, after a sufficient amount of time.}. 
At $K<8$ we observe the same  power-dominated regime described above, and the evolution rapidly converges to a topology optimized for power and characterized by extreme inequality. Wen $8<K<12$ we have a soft power regime where motivated frustration can at least win points over power and the system converges to a structure of lower power and lower Gini indices (in the range 0.7-0.9, observed in USA and UK).

However, as initiative $K$ increases above ${K=12}$ we enter a cyclical behavior. The network {\it does not converge} to a stable society anymore, but instead power and Gini index oscillate around  stable values, which decrease sharply as $K$ increases. In this regime the Gini index spans a range between a minimum of  $\sim 0.1$ and a maximum of slightly more than 0.6 (in this range falls Japan with a  Gini   of  0.55). Interestingly, as $K$ increases so does the amplitude of oscillations. However this window of oscillation abruptly collapses  at $K\sim16$. For $K>16$ the market  converges rapidly again, but now to a regime of self-organized equality, where the Gini index is very small ($\sim 0.1$), corresponding to a random network.

In the cyclical regime, Fig.~4 tells  a tale. A maximum in power corresponds  to {\it a self-organizazion into very distinct lower, middle and upper classes} with respect to opportunities and thus wealth. We remark that in no way this structure is built into the assumptions, it is an emergent structure of the society. These classes represent interdependent~\cite{gao2012networks} networks, each of them resembling a random network, for fairness {\it within} the social class. Looking at $P(z'|z)$ in Fig.~4 we notice that the upper class draws wealth from the middle class, which draws wealth from the lower class. 

Motivated frustration works first through the middle class slowly bringing down the upper class, and thus  improving equality among the top owners, which corresponds to change in shape of the Lorenz curve although with little improvement in the Gini index. As the middle class  broadens, it finally merges with the remnants of an upper class whose wealth has been eroded. After that, quite suddenly, the two remaining classes coalesce in what seems like a single, remarkably equal random network ( $G_w\simeq 0.1$). Yet the society is now so fair that nobody is particularly more frustrated than the others:  the black swans of richness (which appear because of infrequent fluctuations, and are not connected between themselves but draw wealth from the larger, equalized class) can finally win against everybody else. An upper class begins reforming, although it is so tiny that the Gini index is left unchanged. At that point, at lower values of initiative $K$, almost immediately  a middle class appears, the society returns to being suddenly polarized,  and the cycle restarts: the ``time of equality'' is not long lived. But for initiative $K$ close to transition (see $K=16$ in Fig.~4) the time of equality stretches as long as the time of inequality, while an upper class  forms very slowly; after it reaches a critical mass, a middle class forms, bringing back in the time of inequality typical of the 3-class system. 

In our framework, it appears that the birth of an upper class promotes a middle class, and thus indirectly pushes the lower class into poverty. However the middle class works to bring down the upper class. 
Indeed the lower class has frustration but no power. The upper class is powerful yet satiated. It is the middle class, with its mixture of both sufficient power and frustration, that on one hand impoverishes the lower class,  
 on the other brings down the upper class to improve equality among the top owners.

\section{Conclusions}

We have presented a minimal model to describe wealth allocation, alternative to the accepted frameworks based on generational social stratification, to conceptualize the more dynamical contemporary societies. 
Within our model, complete deregulation on wealth transfer brings in  savage inequality, efficiently dispossessing  more than 50\% of the population.
Regulation of a static market, expressed here  through limitation on wealth  transfer, considerably ameliorates the situation, eradicating the  dispossessed, and leading to remarkably fair Gini indices for personal satisfaction, even though the collective equality might differ. 

Driven by the interplay of power and personal frustration/satisfaction, the coevolution of a non-static market does not  converge to just any Gini index value. Depending on the ratio of initiative vs. power, it returns either more or less ameliorated inequality (Gini~$>0.7$) or the extreme of almost complete equality (Gini~$=0.1$). In the middle lies a cyclical regime  of oscillating equality, characterized by the emergence of three distinguishable classes, upper, middle and lower, whose mutual ``3-body'' interaction drives the cyclicity. 
There, periodically, in long times of relative inequality  the middle class works patiently and relentlessly to rise up, to bring down the upper class, and to merge with it. However, when a single egalitarian class forms for a brief time, it is soon disrupted by the appearance of the black swans of richness whose power, now competing against unfrustrated and thus demotivated agents of an egalitarian class, wins easily. A new time of inequality is brought in as a new middle class emerges with the rise of the upper class.  
 
It seems that equality can be improved either by social engineering of a static, proper social topology or more realistically by dynamic, emergent reshaping of the market via sufficient individual initiative: this involves  action on personal frustration, but also individual resistance  to power-moves. That, however, implies universal competence: in our model, the agent needs to know who are its partners, what is their wealth, their relative satisfaction or frustration. 

Even so, however, equality is not stable, as the disappearance of frustration removes its fundamental promoter. Perhaps a key element in preventing the cyclical return of inequality would be {\it memory},  absent from our framework. But  is it present in society?

Finally, one might discuss whether a gradient of wealth inequality impedes or promotes  growth, a most relevant issue which we will explore  in future work, by allowing for wealth creation within our framework. 





\subsection{Author Contributions}
BM conducted the numerical analysis, data interpretation and derived the mean field formula, contributed to the development of the methods, and contributed to the manuscript. CN developed the concept of the study, the early  analytical studies, contributed to the development of the methods, data interpretation and drafting of the manuscript.  CN and AN   supervised the work.

\begin{acknowledgments}
This work was carried out under the auspices
of the NNSA of the U.S. DoE at LANL under Contract No. DE-AC52-06NA25396. BM thanks CNLS for hospitality. CN  thanks L. Bettencourt and R. Albert for discussions. 
\end{acknowledgments}

\bibliography{library.bib}{}

\end{document}